\documentclass[aps]{revtex4}

\usepackage{graphics}
\usepackage{epsfig}
\usepackage{amsmath}

\def\ov#1{\overline{#1}}

\def\vb#1{\mbox{\boldmath$#1$}}
\def\pd#1#2{\frac{\partial #1}{\partial #2}}

\def\wh#1{\widehat{#1}}
\def\bdot{\,\vb{\cdot}\,}
\def\btimes{\,\vb{\times}\,}

\def\bhat{\wh{{\sf b}}}

\def\cal#1{\mathcal{#1}}

\def\eq#1{\eqref{eq:#1}}

\newcommand{\bc}{\begin{center}}
\newcommand{\ec}{\end{center}}
\newcommand{\bt}{\begin{tabbing}}
\newcommand{\et}{\end{tabbing}} 
\newcommand{\be}{\begin{eqnarray*}}
\newcommand{\ee}{\end{eqnarray*}}
\newcommand{\bs}{\begin{slide}}
\newcommand{\es}{\end{slide}}

\begin{document}

\title{Noether derivation of exact conservation laws for dissipationless reduced-fluid models}

\author{A.~J.~Brizard}
\affiliation{Department of Chemistry and Physics \\ Saint Michael's College, Colchester, VT 05439, USA}

\begin{abstract}
The energy-momentum conservation laws for general reduced-fluid (e.g., gyrofluid) models are derived by Noether method from a general reduced variational principle. The reduced canonical energy-momentum tensor (which is explicitly asymmetric and has the Minkowski form) exhibits polarization and magnetization effects associated with dynamical reduction. In particular, the asymmetry in the reduced canonical momentum-stress tensor produces a non-vanishing reduced intrinsic torque that can drive spontaneous toroidal rotation in axisymmetric tokamak plasmas.
\end{abstract}

\begin{flushright}
September 22, 2010 \\
\end{flushright}


\maketitle

\section{Introduction}

Nonlinear reduced-fluid models play an important role in our understanding of the complex dynamical behavior of strongly magnetized plasmas. These nonlinear reduced-fluid models, in which fast time scales such as the compressional Alfv\'{e}n time scale have been asymptotically removed, include the reduced magnetohydrodynamic equations \cite{RMHD_1,RMHD_2,RMHD_3}, the reduced Braginskii equations \cite{RBrag_1,RBrag_2}, the nonlinear gyrofluid equations \cite{Brizard_92,Brizard_Hahm}, and several truncated reduced-fluid models (such as the Hasegawa-Mima equation \cite{HM,Hazeltine_RMHD_HM} and the Hasegawa-Wakatani equations \cite{HW}). Because the space-time-scale orderings for these reduced-fluid models are compatible with the nonlinear gyrokinetic space-time-scale orderings \cite{Brizard_Hahm}, they provide a very useful complementary set of equations that yield simpler interpretations of low-frequency turbulent plasma dynamics in realistic magnetic geometries.

The role of plasma flows in self-regulating anomalous transport processes in turbulent axisymmetric magnetized plasmas has been intensively investigated in the past decade. Because a strong coupling has been observed \cite{Tala_etal,Wang_etal} between toroidal angular-momentum transport and energy transport in such plasmas, it is natural to investigate the link between these two global conservation laws through an application of the Noether method on a suitable Lagrangian density. Axixymmetric tokamak plasmas have also been observed to undergo spontaneous toroidal rotation in the absence of external torque \cite{sp_rot_1,sp_rot_2,sp_rot_3,sp_rot_4,sp_rot_5}. Several attempts have been taken to explain this surprising phenomena \cite{sp_rot_6,sp_rot_7,sp_rot_8,sp_rot_9,sp_rot_10,sp_rot_11}. In the present work, we apply the Noether method on a general reduced Lagrangian density to identify an intrinsic torque generated by the dynamical reduction, which introduces a decoupling between the reduced-fluid momentum and the reduced-fluid velocity as well as reduced polarization and magnetization effects.

The remainder of this paper is organized as follows. In Sec.~\ref{sec:variational}, we present the variational formulation of general reduced-fluid equations that exhibit polarization and magnetization effects in the reduced Maxwell equations as well as a reduced ponderomotive force in the reduced-fluid momentum equation. In Sec.~\ref{sec:energy_mom}, we present the Noether derivation of the reduced energy-momentum conservation laws that are preserved by the reduced-fluid equations. Explicit proofs are also presented that identify the energy-momentum fluxes as well as the energy-momentum transfer terms. The reduced toroidal angular-momentum conservation law is presented in Sec.~\ref{sec:ang_mom}, where the source of reduced intrinsic angular-momentum is identified as the reduced intrinsic torque due to the antisymmetry of the reduced momentum-stress tensor. We show how this asymmetry is explicitly due to the decoupling of the reduced-fuid momentum and the reduced-fluid velocity as well as the polarization and magnetization effects associated with the dynamical reduction. We summarize our work in Sec.~\ref{sec:summary} and discuss future work.

\section{\label{sec:variational}General Variational Formulation}

The general variational formulation of nonlinear dissipationless reduced-fluid models describing nonlinear turbulent dynamics of strongly magnetized plasmas is expressed in terms of a Lagrangian density ${\cal L}(\psi^{\alpha})$ that is a function of the multi-component field
\begin{equation}
\psi^{\alpha} \;\equiv\; (\Phi, {\bf A}, {\bf E}, {\bf B}; n, {\bf u}, p_{\|}, p_{\bot}),
\label{eq:psi_def}
\end{equation} 
where the electromagnetic fields $({\bf E},{\bf B})$ are defined in terms of the electromagnetic potentials $(\Phi,{\bf A})$ as
\begin{equation}
{\bf E} \;\equiv\; -\,\nabla\Phi \;-\; c^{-1}\partial{\bf A}/\partial t \;\;\;{\rm and}\;\;\; {\bf B} \;\equiv\; \nabla\btimes{\bf A}
\label{eq:EB_phiA}
\end{equation} 
and the reduced-fluid moments $(n,{\bf u},p_{\|},p_{\bot})$ are used for each plasma-particle species (with mass $m$ and charge $q$). Gauge invariance requires that the potentials $(\Phi, {\bf A})$ should not appear in the final form of the reduced dynamical equations as well as their associated conservation laws.

We note that the Lagrangian formalism does not accommodate higher-order fluid moments (e.g., heat fluxes) and, therefore, the issue of fluid closures is completely ignored (unless these higher-order moments are expressed in terms of lower-order moments). These higher-order moments, as well as dissipative effects, can thus be added after the dissipationless reduced-fluid equations are derived by variational method (although a variational procedure 
\cite{Peng,Brizard_2005} can be  used to include heat fluxes in the pressure evolution equations). The emphasis of the present work is therefore on nonlinear physics within a Lagrangian formulation of reduced-fluid dynamics.

\subsection{Reduced-fluid Lagrangian}

We now introduce the general Lagrangian density
\begin{equation}
{\cal L}(\psi^{\alpha}) \;\equiv\; {\cal L}_{\rm M}({\bf E}, {\bf B}) \;+\; {\cal L}_{\Psi}(\Phi, {\bf A}; n, {\bf u}) \;+\; {\cal L}_{\rm F}(n, {\bf u}, p_{\|}, p_{\bot}; {\bf E}, {\bf B}), 
\label{eq:L_F_def}
\end{equation}
defined as the sum of the electromagnetic Lagrangian density
\begin{equation}
{\cal L}_{\rm M}({\bf E}, {\bf B}) \;\equiv\; \frac{1}{8\pi} \left( |{\bf E}|^{2} \;-\frac{}{} |{\bf B}|^{2} \right),
\label{eq:Lag_M}
\end{equation}
the gauge-dependent plasma-electromagnetic interaction Lagrangian density
\begin{equation}
{\cal L}_{\Psi}(\Phi, {\bf A}; n, {\bf u}) \;\equiv\; -\;\sum\;q\,n\;\left( \Phi \;-\; {\bf A}\bdot\frac{{\bf u}}{c} \right) \;\equiv\; 
-\;\sum\;q\,n\;\Psi,
\label{eq:Lag_Psi}
\end{equation}
where $\sum$ denotes a sum over particle species, and the reduced-fluid Lagrangian density ${\cal L}_{\rm F}$, which depends on the electromagnetic field 
$({\bf E},{\bf B})$ only through the process of dynamical reduction \cite{Brizard_Vlasovia}.

Next, we introduce the following partial derivatives of the Lagrangian density \eq{L_F_def}. First, we introduce the electromagnetic-potential derivatives
\begin{equation}
\left( \pd{{\cal L}}{\Phi},\; \pd{{\cal L}}{{\bf A}} \right) \;\equiv\; \left( -\,\varrho,\frac{}{} {\bf J}/c \right),
\label{eq:L_phiA}
\end{equation}
which define the reduced charge density $\varrho \equiv \sum\,qn$ and the reduced current density ${\bf J} \equiv \sum\,qn\,{\bf u}$. The definitions \eq{L_phiA} ensure electromagnetic-gauge invariance, which is connected to the charge conservation law
\begin{equation}
-\;\pd{}{t}\left( \pd{{\cal L}}{\Phi}\right) \;+\; \nabla\bdot\left( c\;\pd{{\cal L}}{{\bf A}} \right) \;=\; \pd{\varrho}{t} \;+\; \nabla\bdot{\bf J} 
\;=\; 0
\label{eq:charge_conserve}
\end{equation}
as follows. Under a gauge transformation $\Phi^{\prime} = \Phi - c^{-1}\partial\chi/\partial t$ and ${\bf A}^{\prime} = {\bf A} + \nabla\chi$, we find 
$\Psi^{\prime} \equiv \Psi - c^{-1}d\chi/dt$, where $d/dt \equiv \partial/\partial t + {\bf u}\bdot\nabla$, so that the Lagrangian \eq{L_F_def} transforms as
\[ {\cal L}^{\prime} \;\equiv\; {\cal L} \;+\; \varrho\;\pd{}{t}\left(\frac{\chi}{c}\right) \;+\; {\bf J}\bdot\nabla\left(\frac{\chi}{c} \right) 
\;\equiv\; {\cal L} \;+\; \pd{}{t}\left(\varrho\;\frac{\chi}{c}\right) \;+\; \nabla\bdot\left({\bf J}\;\frac{\chi}{c} \right), \]
where Eq.~\eq{charge_conserve} was used. Hence, the reduced variational principle
\begin{equation}
\delta\;\int\;{\cal L}(\psi^{\alpha})\;d^{4}x \;=\; 0
\label{eq:RVP}
\end{equation}
is gauge-invariant since $\int {\cal L}^{\prime}\,d^{4}x = \int {\cal L}\,d^{4}x$. We therefore see that gauge invariance and charge conservation are intimately connected within a Lagrangian formulation.

Second, we introduce the electromagnetic-field derivatives \cite{Brizard_Vlasovia,Brizard_NFLR,Brizard_JPCS}
\begin{equation}
\left( \pd{{\cal L}}{{\bf E}},\; \pd{{\cal L}}{{\bf B}}\right) \;\equiv\; \left( \frac{{\bf D}}{4\pi},\; -\,\frac{{\bf H}}{4\pi} \right) \;\equiv\; 
\left( \frac{{\bf E}}{4\pi} \;+\; {\bf P},\; -\,\frac{{\bf B}}{4\pi} \;+\; {\bf M} \right).
\label{eq:L_EB}
\end{equation}
which define the reduced polarization and magnetization
\begin{equation}
\left( {\bf P},\; {\bf M} \right) \;\equiv\; \left( \pd{{\cal L}_{\rm F}}{{\bf E}},\; \pd{{\cal L}_{\rm F}}{{\bf B}} \right).
\label{eq:pol_mag}
\end{equation} 
These definitions correspond to polarization and magnetization effects associated with reduced electric and magnetic dipole moments \cite{Brizard_Vlasovia,Brizard_JPCS}. Higher-order multipole contributions require the reduced-fluid Lagrangian density to depend on gradients of the electromagnetic fields (which are not considered in the present paper). We note that gyrofluid models \cite{Brizard_92} and gyrokinetic models \cite{Brizard_Hahm} implicitly contain all multipole polarization and magnetization contributions whenever finite-Larmor-radius effects are retained to all orders.

Third, we introduce the reduced-fluid derivatives
\begin{equation}
\left. \begin{array}{rcl} 
K & \equiv & \partial{\cal L}_{\rm F}/\partial n \\
 &  & \\
{\bf p} & \equiv & n^{-1}\;\partial{\cal L}_{\rm F}/\partial{\bf u}
\end{array} \right\}, 
\label{eq:L_nu}
\end{equation}
which define the reduced kinetic energy $K$ and the reduced kinetic momentum ${\bf p}$. Lastly, we introduce the symmetric reduced pressure tensor
\begin{equation}
{\sf P}_{*} \;\equiv\; -\;2\,p_{\|}\,\pd{{\cal L}_{\rm F}}{p_{\|}}\;\bhat\bhat \;-\; p_{\bot}\,\pd{{\cal L}_{\rm F}}{p_{\bot}}\;
\left({\bf I} - \bhat\bhat\right) \;\equiv\; p_{\|}\,\gamma_{\|}\;\bhat\bhat \;+\; p_{\bot}\,\gamma_{\bot}\;\left({\bf I} - \bhat\bhat\right).
\label{eq:L_p}
\end{equation}
We note that standard finite-Larmor-radius (FLR) corrections are automatically included in the reduced pressure tensor \eq{L_p} through the term 
$\gamma_{\bot} \neq 1$ \cite{SSB}.

reduced-fluid models derived from the variational principle \eq{RVP} will be distinguished by the physical effects contained in the fluid Lagrangian density \eq{L_F_def} through the fields defined in Eqs.~(\ref{eq:L_phiA})-(\ref{eq:L_p}).

\subsection{Constraint Equations}

The variational principle (\ref{eq:RVP}) used to derive the nonlinear dissipationless dynamical equations relies on certain constraint equations that must be preserved by the field variations $\delta\psi^{\alpha}$. First, the electromagnetic variations
\begin{equation}
\delta{\bf E} \;\equiv\; -\;\nabla\delta\Phi \;-\; c^{-1}\partial\delta{\bf A}/\partial t \;\;\;{\rm and}\;\;\; \delta{\bf B} \;\equiv\; \nabla\btimes
\delta{\bf A}
\label{eq:delta_EB}
\end{equation} 
preserve the electromagnetic constraints
\begin{equation}
\nabla\btimes{\bf E} \;=\; -\,c^{-1}\partial{\bf B}/\partial t \;\;\;{\rm and}\;\;\; \nabla\bdot{\bf B} \;=\; 0.
\label{eq:Maxwell_00}
\end{equation}

For the reduced-fluid moments $(n,{\bf u},p_{\|},p_{\bot})$ associated with each particle species, the constraint equations are the continuity equation (associated with the conservation of particle number)
\begin{equation}
\pd{n}{t} \;=\; -\;\nabla\bdot\left(n\frac{}{}{\bf u} \right),
\label{eq:n_eq}
\end{equation}
and the pressure equations
\begin{eqnarray}
\pd{p_{\|}}{t} & = & -\;\nabla\bdot\left(p_{\|}\frac{}{}{\bf u} \right) \;-\; 2\,p_{\|}\;\bhat\bhat:\nabla{\bf u}, \label{eq:ppar_eq} \\
\pd{p_{\bot}}{t} & = & -\;\nabla\bdot\left(p_{\bot}\frac{}{}{\bf u} \right) \;-\; p_{\bot}\;\left({\bf I} - \bhat\bhat\right):\nabla{\bf u}, 
\label{eq:pper_eq}
\end{eqnarray}
where the Chew-Goldberger-Low (CGL) pressure tensor is \cite{CGL}
\begin{equation}
{\sf P} \;\equiv\; p_{\|}\;\bhat\,\bhat \;+\; p_{\bot}\;\left( {\bf I} \;-\; \bhat\,\bhat\right).
\label{eq:P_CGL}
\end{equation}
Recall that the CGL pressure equations \eq{ppar_eq}-\eq{pper_eq} arise from the adiabatic conservation laws \cite{CGL,RR}
\begin{eqnarray}
\frac{d}{dt} \left( \frac{p_{\bot}}{n\,B} \right) & = & 0, \label{eq:p_perp} \\
\frac{d}{dt} \left( \frac{p_{\|}\,B^{2}}{n^{3}} \right) & = & 0, \label{eq:p_par} 
\end{eqnarray}
associated with the adiabatic invariance of the gyro-action (or magnetic moment) and the bounce-action, respectively, where $\partial B/\partial t 
\equiv \bhat\bdot\nabla\btimes({\bf u}\btimes{\bf B})$ is used in Eqs.~\eq{p_perp}-\eq{p_par}. The variations $(\delta n, \delta{\bf u}, 
\delta p_{\bot}, \delta p_{\|})$ that preserve the constraint equations \eq{n_eq} and \eq{p_perp}-\eq{p_par} are
\begin{equation}
\left. \begin{array}{rcl}
\delta n & = & -\;\nabla\bdot\left(n\;\vb{\xi}\right) \\
 && \\
\delta{\bf u} & = & \left( \partial/\partial t \;+\; {\bf u}\bdot\nabla\right)\vb{\xi} \;-\; \vb{\xi}\bdot\nabla{\bf u} \\
 && \\
\delta p_{\bot} & = & -\;\nabla\bdot\left(p_{\bot}\;\vb{\xi}\right) \;-\; p_{\bot}\:({\bf I} - \bhat\bhat):\nabla\vb{\xi} \\
 && \\
\delta p_{\|} & = & -\;\nabla\bdot\left(p_{\|}\;\vb{\xi}\right) \;-\; 2\,p_{\|}\:\bhat\bhat:\nabla\vb{\xi}
\end{array} \right\},
\label{eq:delta_fluid}
\end{equation}
where $\vb{\xi}$ generates a virtual spatial displacement for a fluid element of each particle species and $\delta B \equiv \bhat\bdot\nabla\btimes(\vb{\xi}\btimes{\bf B})$ is used in order to satisfy the CGL adiabatic conservation laws \eq{p_perp}-\eq{p_par}. Note here that magnetic field ${\bf B} \equiv B\,\bhat$ considered in Eq.~\eq{delta_fluid} involves the general magnetic field and not just the background time-independent magnetic field used in our previous work \cite{Brizard_NFLR}.

\subsection{Euler-Poincar\'{e} Equations}

Using the variations \eq{delta_EB} and \eq{delta_fluid}, the variation of the Lagrangian density ${\cal L}(\psi^{\alpha})$ for a general reduced-fluid model is expressed in terms of the variation fields $(\delta\Phi, \delta{\bf A}, \vb{\xi})$ as
\begin{eqnarray}
\delta{\cal L} & = & \pd{\Lambda}{t} \;+\; \nabla\bdot\vb{\Gamma} \;+\; \delta\Phi \left( \pd{{\cal L}}{\Phi} \;+\; \nabla\bdot\pd{{\cal L}}{{\bf E}} 
\right) \;+\; \delta{\bf A}\bdot\left( \pd{{\cal L}}{{\bf A}} \;+\; \frac{1}{c}\,\pd{}{t}\,\pd{{\cal L}}{{\bf E}} \;+\; \nabla\btimes
\pd{{\cal L}}{{\bf B}} \right) \nonumber \\
 &  &-\; \sum\;\vb{\xi}\bdot\left[\; \pd{}{t}\,\pd{{\cal L}}{{\bf u}} \;+\; \nabla\bdot\left( {\bf u}\,\pd{{\cal L}}{{\bf u}} \right) \;+\; \nabla{\bf u}
\bdot\pd{{\cal L}}{{\bf u}} \;-\; \left( \eta^{a}\;\nabla\pd{{\cal L}}{\eta^{a}} \right) \;+\; \nabla\bdot{\sf P}_{*} \;\right],
\label{eq:Lag_var}
\end{eqnarray}
where $\eta^{a} = (n, p_{\|}, p_{\bot})$ combines the density and pressure fluid moments (summation over repeated indices is implied), and the Noether density $\Lambda$ and flux $\vb{\Gamma}$ are
\begin{eqnarray}
\Lambda & = & \sum\;\vb{\xi}\bdot\pd{{\cal L}}{{\bf u}} \;-\; \frac{1}{c}\,\delta{\bf A}\bdot\pd{{\cal L}}{{\bf E}}, \label{eq:Lambda} \\
\vb{\Gamma} & = & \sum\; \left[\; {\bf u} \left( \vb{\xi}\bdot\pd{{\cal L}}{{\bf u}} \right) \;-\; \vb{\xi} 
\left(\eta^{a}\,\pd{{\cal L}}{\eta^{a}} \right) \;+\; {\sf P}_{*}\bdot\vb{\xi} \;\right] \;-\; \delta\Phi\;\pd{{\cal L}}{{\bf E}} \;+\; \delta{\bf A}\btimes\pd{{\cal L}}{{\bf B}}. 
\label{eq:Gamma}
\end{eqnarray}
Note that the Noether terms $\partial\Lambda/\partial t + \nabla\bdot\vb{\Gamma}$ in Eq.~\eq{Lag_var} do not contribute to the variational principle 
(\ref{eq:RVP}) since they appear as a space-time divergence.

For arbitrary variations $(\delta \Phi, \delta {\bf A}, \vb{\xi})$, the variational principle \eq{RVP} yields the Euler-Poincar\'{e} equations 
\begin{eqnarray}
0 & = & \pd{{\cal L}}{\Phi} \;+\; \nabla\bdot\pd{{\cal L}}{{\bf E}}, 
\label{eq:Phi_EP} \\
0 & = & \pd{{\cal L}}{{\bf A}} \;+\; \frac{1}{c}\,\pd{}{t}\,\left(\pd{{\cal L}}{{\bf E}}\right) \;+\; \nabla\btimes\pd{{\cal L}}{{\bf B}}, 
\label{eq:A_EP} \\
0 & = & \pd{}{t}\,\left(\pd{{\cal L}}{{\bf u}}\right) \;+\; \nabla\bdot\left( {\bf u}\,\pd{{\cal L}}{{\bf u}} \right) \;+\; \nabla{\bf u}
\bdot\pd{{\cal L}}{{\bf u}} \;-\; \left( \eta^{a}\;\nabla\pd{{\cal L}}{\eta^{a}} \right) \;+\; \nabla\bdot{\sf P}_{*}. 
\label{eq:u_EP}
\end{eqnarray}
The first two equations yield the reduced Maxwell equations
\begin{eqnarray}
\nabla\bdot{\bf D} & = & 4\pi\;\varrho, \label{eq:D_eq} \\
\nabla\btimes{\bf H} \;-\; \frac{1}{c}\;\pd{{\bf D}}{t} & = & \frac{4\pi}{c}\;{\bf J}, \label{eq:H_eq}
\end{eqnarray}
where we used the definitions \eq{L_phiA}-\eq{L_EB}. These reduced Maxwell equations can also be expressed in terms of the electromagnetic fields 
$({\bf E},{\bf B})$ as
\begin{eqnarray}
\nabla\bdot{\bf E} & = & 4\pi\;\left( \varrho \;-\frac{}{} \nabla\bdot{\bf P}\right), \label{eq:DE_eq} \\
\nabla\btimes{\bf B} \;-\; \frac{1}{c}\;\pd{{\bf E}}{t} & = & \frac{4\pi}{c}\;\left( {\bf J} \;+\; \pd{{\bf P}}{t} \;+\; c\;\nabla\btimes{\bf M}\right), \label{eq:HB_eq}
\end{eqnarray}
where $\varrho_{\rm pol} \equiv -\,\nabla\bdot{\bf P}$ denotes the polarization density, ${\bf J}_{\rm pol} \equiv \partial{\bf P}/\partial t$ denotes the polarization current, and ${\bf J}_{\rm mag} \equiv c\,\nabla\btimes{\bf M}$ denotes the magnetization current.

Equation \eq{u_EP} yields the time evolution equation for the fluid velocity ${\bf u}$, subject to the constraint equations \eq{n_eq}-\eq{pper_eq}. Using the definitions \eq{L_nu}-\eq{L_p}, Eq.~\eq{u_EP} becomes
\begin{equation}
n \left( \pd{}{t} \;+\; {\bf u}\bdot\nabla\right){\bf p} \;=\; qn\; \left( {\bf E} \;+\; \frac{{\bf u}}{c}\btimes{\bf B} \right) \;+\; n \left( \nabla K
\;-\; \nabla{\bf u}\bdot{\bf p} \right) \;-\; \left( p_{\bot}\;\nabla\gamma_{\bot} \;+\; \frac{p_{\|}}{2}\;\nabla\gamma_{\|} \;+\; \nabla\bdot{\sf P}_{*} \right),
\label{eq:pmom_eq}
\end{equation}
where the first term on the right side is the Lorentz force on the charged reduced-fluid, the terms involving $\nabla K$ and $\nabla{\bf u}\bdot{\bf p}$ involve the reduced kinetic energy and the reduced kinetic momentum \eq{L_nu}, while the remaining terms involve generalizations of the CGL pressure-tensor force (with $\gamma_{\bot}$, $\gamma_{\|} \neq 1$).

For a regular particle-fluid model, where the particle-fluid Lagrangian is ${\cal L}_{\rm pF} \equiv mn\,|{\bf u}|^{2}/2 - {\cal P}$, where ${\cal P} \equiv \frac{1}{2}\,{\rm Tr}({\sf P}) = p_{\bot} + p_{\|}/2$ and reduced polarization-magnetization effects are absent, the particle-fluid kinetic momentum is ${\bf p} = m\,{\bf u}$ and the particle-fluid kinetic energy is 
$K = m\,|{\bf u}|^{2}/2$, so that $\nabla K = \nabla{\bf u}\bdot{\bf p}$. Next, the particle-fluid pressure derivatives are $\gamma_{\|} = 1 = 
\gamma_{\bot}$ so that the reduced pressure tensor ${\sf P}_{*} \equiv {\sf P}$ is simply given by the CGL pressure tensor (\ref{eq:P_CGL}). Equation 
(\ref{eq:pmom_eq}) therefore becomes the particle-fluid momentum equation
\[ \frac{d{\bf p}}{dt} \;=\; q\;\left( {\bf E} \;+\; \frac{{\bf u}}{c}\btimes{\bf B} \right) \;-\; n^{-1}\;\nabla\bdot{\sf P}, \] 
where the CGL pressure-tensor force
\[ \nabla\bdot{\sf P} \;=\; \nabla p_{\bot} \;+\; p_{\Delta}\;\bhat\bdot\nabla\bhat \;+\; \left[\nabla\bdot\left( p_{\Delta}\;\bhat\right)\right]\;
\bhat \]
is expressed in terms of the pressure anisotropy $p_{\Delta} \equiv p_{\|} - p_{\bot}$.

\section{\label{sec:energy_mom}Reduced Energy-Momentum Conservation Laws}

The purpose of the present Section is to derive the reduced energy-momentum conservation laws by applying the Noether method on the reduced Lagrangian density \eq{L_F_def}. We also present explicit proofs of energy-momentum conservation for these reduced-fluid models in order to uncover the energy-momentum transfer processes. The reduced toroidal angular momentum conservation law will be considered in the next Section.

\subsection{Noether method}

After substituting the Euler-Poincar\'{e} equations \eq{Phi_EP}-\eq{u_EP} into the variation \eq{Lag_var}, we obtain the Noether equation
\begin{equation}
\delta{\cal L} \;=\; \pd{\Lambda}{t} \;+\; \nabla\bdot\vb{\Gamma}.
\label{eq:Noether_eq}
\end{equation}
The energy-momentum conservation laws are constructed from the Noether equation \eq{Noether_eq} based on symmetries of the Lagrangian density
\eq{L_F_def} with respect to space-time translations generated by $(c\,\delta t, \delta{\bf x})$. 

First, we introduce the variations induced by the space-time translations
\begin{equation}
\left. \begin{array}{rcl}
\vb{\xi} & = & \delta{\bf x} \;-\; {\bf u}\;\delta t \\
 & & \\
\delta{\cal L} & = & -\,\delta t\;(\partial{\cal L}/\partial t - \partial^{\prime}{\cal L}/\partial t) \;-\; \delta{\bf x}\bdot(\nabla{\cal L} - 
\nabla^{\prime}{\cal L}) \\
 & & \\
\delta\Phi & = & \delta{\bf x}\bdot{\bf E} \;-\; c^{-1}\partial\delta\chi/\partial t \\
 & & \\
\delta{\bf A} & = & \delta{\bf x}\btimes{\bf B} \;+\; c\,\delta t\;{\bf E} \;+\; \nabla\delta\chi
\end{array} \right\},
\label{eq:var_xt}
\end{equation}
where the gauge-dependent term
\begin{equation}
\delta\chi \;\equiv\; c\,\delta t\;\Phi \;-\; \delta{\bf x}\bdot{\bf A}
\label{eq:delta_chi}
\end{equation} 
will be removed below and
\[ \left( \nabla^{\prime}{\cal L}, \frac{\partial^{\prime}{\cal L}}{\partial t}\right) \;\equiv\; \left( \nabla{\cal L} \;-\; \nabla\psi^{\alpha}\;
\pd{{\cal L}}{\psi^{\alpha}},\; \pd{{\cal L}}{t} \;-\; \pd{\psi^{\alpha}}{t}\;\pd{{\cal L}}{\psi^{\alpha}} \right) \]
denote space-time derivatives with the multi-component field $\psi^{\alpha}$ held fixed. By substituting the variations \eq{var_xt}, we obtain the Noether components
\begin{eqnarray}
\Lambda & = & \sum\;\left( \delta{\bf x} \;-\frac{}{} {\bf u}\;\delta t \right)\bdot\pd{{\cal L}}{{\bf u}} \;-\; \frac{1}{c} \left( \delta{\bf x}\btimes
{\bf B} \;+\; c\,\delta t\;{\bf E} \;+\frac{}{} \nabla\delta\chi \right)\bdot\pd{{\cal L}}{{\bf E}}, \label{eq:Lambda_xt} \\
\vb{\Gamma} & = & \sum\;\left\{\; {\bf u} \left[ \left( \delta{\bf x} \;-\frac{}{} {\bf u}\;\delta t \right)\bdot\pd{{\cal L}}{{\bf u}} \right] \;+\;
\left[ {\sf P}_{*} \;-\; \left( \eta^{a}\,\pd{{\cal L}}{\eta^{a}} \right) {\bf I} \right]\bdot \left( \delta{\bf x} \;-\frac{}{} {\bf u}\;\delta t 
\right)\;\right\} \nonumber \\
 &  &-\; \left( \delta{\bf x}\bdot{\bf E} \;-\; \frac{1}{c}\,\pd{\delta\chi}{t}\right)\;\pd{{\cal L}}{{\bf E}} \;+\; \left( \delta{\bf x}\btimes
{\bf B} \;+\; c\,\delta t\;{\bf E} \;+\frac{}{} \nabla\delta\chi \right)\btimes\pd{{\cal L}}{{\bf B}}. \label{eq:Gamma_xt}
\end{eqnarray}

The gauge-dependent terms in Eqs.~\eq{Lambda_xt}-\eq{Gamma_xt} can now be removed as follows. First, we note that the right side of the Noether equation \eq{Noether_eq} is invariant under the transformation \cite{Belinfante,McLennan}
\begin{equation}
\left. \begin{array}{rcl}
\ov{\Lambda} & \equiv & \Lambda \;+\; \nabla\bdot{\bf q} \\
 &  & \\
\ov{\vb{\Gamma}} & \equiv & \vb{\Gamma} \;-\; \partial{\bf q}/\partial t \;-\; c\,\nabla\btimes{\bf m}
\end{array} \right\},
\label{eq:Noether_gauge}
\end{equation}
where ${\bf q}$ and ${\bf m}$ are arbitrary vector fields. For example, by substituting the following gauge-dependent vector fields in 
Eq.~\eq{Noether_gauge},
\begin{equation}
{\bf q} \;\equiv\; \frac{\delta\chi}{c}\;\pd{{\cal L}}{{\bf E}} \;\;{\rm and}\;\; {\bf m} \;\equiv\; \frac{\delta\chi}{c}\;\pd{{\cal L}}{{\bf B}},
\label{eq:qm_gauge}
\end{equation}
we obtain the new Noether components
\begin{eqnarray}
\ov{\Lambda} & \equiv & \sum\;\left[ \left( \delta{\bf x} \;-\frac{}{} {\bf u}\;\delta t \right)\bdot n\,{\bf p} \right] 
\;-\; \frac{1}{c} \left( \delta{\bf x}\btimes{\bf B} \;+\frac{}{} c\,\delta t\;{\bf E}\right)\bdot\frac{{\bf D}}{4\pi} \;-\; {\cal L}_{\Psi}\;\delta t, \label{eq:Lambda_bar} \\
\ov{\vb{\Gamma}} & \equiv & \sum\;\left\{\; \left[ {\sf P}_{*} \;+\; n\,{\bf u}\,{\bf p} \;-\; \left( \eta^{a}\;\pd{{\cal L}_{\rm F}}{\eta^{a}} \right) 
{\bf I} \right]\bdot \left( \delta{\bf x} \;-\frac{}{} {\bf u}\;\delta t \right) \right\} \;-\; {\cal L}_{\Psi}\;\delta{\bf x} \nonumber \\
 &  &-\; \left[ {\bf D}\,{\bf E} \;+\; {\bf B}\,{\bf H} \;-\; \left({\bf B}\bdot{\bf H}\right)\frac{}{}{\bf I}\right]\bdot\frac{\delta{\bf x}}{4\pi} 
\;-\; \delta t\;\left( \frac{c}{4\pi}\;{\bf E}\btimes{\bf H} \right).
\label{eq:Gamma_bar}
\end{eqnarray}
In Eqs.~\eq{Lambda_bar}-\eq{Gamma_bar}, the gauge-dependent interaction Lagrangian density ${\cal L}_{\Psi} \equiv -\;\varrho\,\Phi + {\bf A}\bdot{\bf J}/c$ will be cancelled by the specific form \eq{L_F_def} of the Lagrangian density ${\cal L}$ through the term $\delta{\cal L}_{\Psi} \equiv -\,\delta t\,\partial{\cal L}_{\Psi}/\partial t - \delta{\bf x}\bdot\nabla{\cal L}_{\Psi}$ on the left side of the Noether equation \eq{Noether_eq}.

Now that all gauge dependence has been explicitly removed from our Lagrangian formulation of reduced-fluid dynamics, we redefine the Lagrangian density \eq{L_F_def} as $\ov{{\cal L}} \equiv {\cal L}_{\rm M} + {\cal L}_{\rm F}$ as the sum of the Maxwell Lagrangian density and the reduced-fluid density and the Noether equation \eq{Noether_eq} becomes
\begin{equation}
\delta\ov{{\cal L}} \;\equiv\; \pd{\ov{\Lambda}}{t} \;+\; \nabla\bdot\ov{\vb{\Gamma}},
\label{eq:Noether_bar}
\end{equation}
where the right side is still subject to an additional transformation \eq{Noether_gauge}.

A standard perturbative approach to investigating turbulent dynamics in magnetized plasmas is to separate the time-independent nonuniform magnetic field ${\bf B}_{0} \equiv \nabla\btimes{\bf A}_{0}$ from the time-dependent dynamical plasma-electrodynamic fields. Based on this separation, we can substitute 
\[ \frac{\partial^{\prime}\ov{{\cal L}}}{\partial t} \;=\; \pd{{\bf B}_{0}}{t}\bdot\pd{\ov{{\cal L}}}{{\bf B}_{0}} \;\equiv\; 0 \]
in the reduced energy conservation law, and the dynamical plasma-electrodynamic fields are able to exchange energy among themselves while conserving total (global) energy. 

The nonuniformity of the background magnetic field ${\bf B}_{0}$ implies that the term $\nabla^{\prime}\ov{{\cal L}} \neq 0$ in the momentum conservation law drives the exchange of momentum (including angular momentum) between the background magnetic field and the dynamical plasma-electrodynamic fields \cite{Dewar_77}. The Noether Theorem, however, tells us that the total momentum is conserved in the direction of spatial symmetry of the background magnetic field. Hence, the total toroidal angular momentum is conserved in an axisymmetric magnetized plasma when the intrinsic angular momentum is taken into account (see next Section). 

\subsection{Reduced energy conservation law}

In a time-independent medium, the reduced energy conservation law 
\begin{equation}
\pd{{\cal E}}{t} \;+\; \nabla\bdot{\bf S} \;=\; -\;\frac{\partial^{\prime}\ov{{\cal L}}}{\partial t} \;\equiv\; 0,
\label{eq:energy_t}
\end{equation}
is expressed in terms of the reduced energy density
\begin{equation}
{\cal E} \;\equiv\; \sum\; n\;{\bf p}\bdot{\bf u} \;+\; {\bf E}\bdot{\bf P} \;+\; \frac{1}{8\pi} \left( |{\bf E}|^{2} \;+\; |{\bf B}|^{2} \right) 
\;-\; {\cal L}_{\rm F},
\label{eq:E_def}
\end{equation}
and the reduced energy-density flux
\begin{equation}
{\bf S} \;\equiv\; \sum\;\left[\; {\bf u} \left( n\,{\bf p}\bdot{\bf u} \;-\; \eta^{a}\;\pd{{\cal L}_{\rm F}}{\eta^{a}} \right) \;+\; 
{\sf P}_{*}\bdot{\bf u}\;\right] \;+\; \frac{c}{4\pi}\;{\bf E}\btimes{\bf H}.
\label{eq:S_def}
\end{equation}
The energy conservation law \eq{energy_t} is normally used to benchmark the accuracy of reduced-fluid numerical codes (see Ref.~\cite{Brizard_NFLR} for example). For a regular particle-fluid, the standard expressions for the energy density 
\[ {\cal E}_{\rm p} \;=\; \frac{mn}{2}\,|{\bf u}|^{2} \;+\; {\cal P} \;+\; \frac{1}{8\pi}\; \left( |{\bf E}|^{2} + |{\bf B}|^{2} \right) \]
and the energy-density flux 
\[ {\bf S}_{\rm p} \;=\; {\bf u}\,\left( \frac{mn}{2}\,|{\bf u}|^{2} \;+\; {\cal P} \right) \;+\; {\sf P}\bdot{\bf u} \;+\; \frac{c}{4\pi}\;{\bf E}
\btimes{\bf B} \]
are directly obtained from Eqs.~\eq{E_def}-\eq{S_def}.

To prove the reduced energy conservation law \eq{energy_t}, we begin with the partial time derivative of the energy density \eq{E_def}
\begin{equation} 
\pd{{\cal E}}{t} \;=\; \sum\;\pd{(n{\bf p})}{t}\bdot{\bf u} \;+\; \left( \frac{{\bf E}}{4\pi}\bdot\pd{{\bf D}}{t} \;+\; \frac{{\bf B}}{4\pi}\bdot\pd{{\bf B}}{t} \right) \;+\; \left( \sum\,\pd{{\cal L}_{\rm F}}{{\bf u}}\bdot\pd{{\bf u}}{t} \;+\; \pd{{\cal L}_{\rm F}}{{\bf E}}\bdot\pd{{\bf E}}{t} \;-\; \pd{{\cal L}_{\rm F}}{t} \right).
\label{eq:E_dot}
\end{equation}
Upon using Eqs.~\eq{n_eq} and \eq{pmom_eq}, the kinetic-energy term on the right side of Eq.~\eq{E_dot} becomes
\begin{equation}
\sum\;\pd{(n{\bf p})}{t}\bdot{\bf u} \;=\; {\bf J}\bdot{\bf E} \;-\; \sum \left[\; \nabla\bdot\left( n{\bf u}\frac{}{}{\bf p}\bdot{\bf u}\right) \;-\; \eta^{a}\;{\bf u}\bdot\nabla\pd{{\cal L}_{\rm F}}{\eta^{a}} \;+\; \left(\nabla\bdot{\bf P}_{*}\right)\bdot{\bf u} \;\right].
\label{eq:E_1}
\end{equation}
Next, using Eqs.~\eq{Maxwell_00} and \eq{H_eq}, the electromagnetic-energy terms become
\begin{equation}
\frac{{\bf E}}{4\pi}\bdot\pd{{\bf D}}{t} \;+\; \frac{{\bf B}}{4\pi}\bdot\pd{{\bf B}}{t} \;=\; -\;\nabla\bdot\left( 
\frac{c}{4\pi}\;{\bf E}\btimes{\bf H} \right) \;+\; \pd{{\cal L}_{\rm F}}{{\bf B}}\bdot\pd{{\bf B}}{t} \;-\; {\bf J}\bdot{\bf E},
\label{eq:E_4}
\end{equation}
while using Eqs.~\eq{L_EB}-\eq{L_p} and \eq{n_eq}-\eq{pper_eq}, the last term becomes (using $\partial^{\prime}\ov{{\cal L}}/
\partial t \equiv 0$)
\begin{equation}
\sum\,\pd{{\cal L}_{\rm F}}{{\bf u}}\bdot\pd{{\bf u}}{t} \;+\; \pd{{\cal L}_{\rm F}}{{\bf E}}\bdot\pd{{\bf E}}{t} \;-\; \pd{{\cal L}_{\rm F}}{t} \;=\; \sum\; \left[\; \nabla\bdot\left(\eta^{a}\,{\bf u}\right)\;\pd{{\cal L}_{\rm F}}{\eta^{a}} \;-\; {\sf P}_{*}:\nabla{\bf u} \;\right] \;-\; 
\pd{{\cal L}_{\rm F}}{{\bf B}}\bdot\pd{{\bf B}}{t}.
\label{eq:E_5}
\end{equation}
By combining Eqs.~\eq{E_1}-\eq{E_5} into Eq.~\eq{E_dot}, we obtain the reduced energy conservation law \eq{energy_t}.

\subsection{Reduced momentum conservation law}

In a nonuniform medium, the reduced momentum conservation law \cite{footnote}
\begin{equation}
\pd{\vb{\Pi}^{\rm can}}{t} \;+\; \nabla\bdot{\sf T} \;=\; \nabla^{\prime}\ov{{\cal L}}, 
\label{eq:momentum_t}
\end{equation}
is expressed in terms of the reduced canonical momentum density
\begin{equation}
\vb{\Pi}^{\rm can} \;\equiv\; \sum\; n\;{\bf p} \;+\; \frac{{\bf D}\btimes{\bf B}}{4\pi\,c},
\label{eq:pi_def}
\end{equation}
which has the Minkowski form \cite{Abr_Mink}, and the reduced canonical momentum-stress tensor
\begin{eqnarray}
{\sf T} & \equiv & \sum\;{\sf P}_{*} \;+\; \left[\; \left( {\cal L}_{\rm F} \;-\; \sum\; \eta^{a}\;\pd{{\cal L}_{\rm F}}{\eta^{a}}\right) \;+\; 
\frac{1}{8\pi} \left( |{\bf E}|^{2} \;+\; |{\bf B}|^{2} \right) \;-\; {\bf B}\bdot{\bf M} \;\right]\;{\bf I} \nonumber \\
 &  &+\; \left[\; \sum\;n\;{\bf u}\;{\bf p} \;-\; \frac{1}{4\pi}\; \left( {\bf D}\;{\bf E} \;+\frac{}{} {\bf B}\;{\bf H} \right) \;\right].
\label{eq:T_def}
\end{eqnarray}
While the first two terms in the reduced canonical momentum-stress tensor \eq{T_def} are obviously symmetric, the remaining terms are not in a reduced-fluid model in which $({\bf p}, {\bf D}, {\bf H}) \neq (m{\bf u}, {\bf E}, {\bf B})$. We note that, for a regular particle-fluid, the standard expressions for the canonical momentum density 
\[ \vb{\Pi}_{\rm p}^{\rm can} \;=\; \sum\;mn\,{\bf u} \;+\; \frac{{\bf E}\btimes{\bf B}}{4\pi\;c} \]
and the symmetric canonical momentum-stress tensor 
\[ {\sf T}_{\rm p} \;=\; \sum\;\left( {\sf P} \;+\frac{}{} mn\,{\bf u}\,{\bf u}\right) \;+\; \frac{1}{4\pi} \left[\; \left( |{\bf E}|^{2} \;+\frac{}{} 
|{\bf B}|^{2} \right)\;\frac{{\bf I}}{2} \;-\; \left( {\bf E}\,{\bf E} + {\bf B}\,{\bf B} \right) \;\right] \]
are directly obtained from Eqs.~\eq{pi_def}-\eq{T_def}. It is therefore clear that the asymmetry of the reduced canonical momentum-stress tensor 
\eq{T_def} is due to the dynamical reduction that led to the reduced-fluid Lagrangian density ${\cal L}_{\rm F}$.

We now prove the reduced momentum conservation law \eq{momentum_t} by calculating explicitly the partial time derivative of the reduced canonical momentum density \eq{pi_def}
\begin{equation}
\pd{\vb{\Pi}^{\rm can}}{t} \;=\; \sum\; \pd{(n\,{\bf p})}{t} \;+\; \pd{}{t} \left( \frac{{\bf D}\btimes{\bf B}}{4\pi\,c} \right).
\label{eq:Pi_dot}
\end{equation}
First, from Eq.~\eq{pmom_eq}, we find
\begin{equation}
\pd{(n\,{\bf p})}{t} \;=\; -\;\nabla\bdot\left(n\,{\bf u}\frac{}{}{\bf p}\right) \;+\; qn\; \left( {\bf E} \;+\; \frac{{\bf u}}{c}\btimes{\bf B} \right) 
\;+\; \eta^{a}\;\nabla\pd{{\cal L}_{\rm F}}{\eta^{a}} \;-\; \nabla\bdot{\sf P}_{*} \;-\; \nabla{\bf u}\bdot \pd{{\cal L}_{\rm F}}{{\bf u}}.
\label{eq:np_dot}
\end{equation}
Next, from Eqs.~\eq{Maxwell_00} and \eq{H_eq}, we find
\begin{eqnarray}
\pd{}{t} \left( \frac{{\bf D}\btimes{\bf B}}{4\pi\,c} \right)  & = & \nabla\bdot\left[\; \frac{{\bf B}\,{\bf H}}{4\pi} \;+\; \frac{{\bf D}\,
{\bf E}}{4\pi} \;-\; \left( \frac{{\bf B}\bdot{\bf H}}{4\pi} \right) {\bf I} \;\right] \;-\; \frac{1}{4\pi} \left( \nabla{\bf E}\bdot{\bf D} \;-\frac{}{} \nabla{\bf B}\bdot{\bf H}\right) \nonumber \\
 &  &-\; \sum\; n\,q \left( {\bf E} \;+\; \frac{{\bf u}}{c}\btimes{\bf B} \right).
\label{eq:DxB_dot}
\end{eqnarray}
Lastly, we add Eqs.~\eq{np_dot}-\eq{DxB_dot} with the identity
\begin{eqnarray}
0 & = & \left( \nabla^{\prime}\ov{{\cal L}} \;-\frac{}{} \nabla\ov{{\cal L}}\right) \;+\; \sum \left( \nabla{\bf u}\bdot 
\pd{{\cal L}_{\rm F}}{{\bf u}} \;+\; \nabla\eta^{a}\;\pd{{\cal L}_{\rm F}}{\eta^{a}} \right) \;+\; \nabla{\bf E}\bdot\pd{\ov{{\cal L}}}{{\bf E}} 
\;+\; \nabla{\bf B}\bdot\pd{\ov{{\cal L}}}{{\bf B}} \label{eq:grad_L} \\
 & = & \nabla^{\prime}\ov{{\cal L}} \;-\; \nabla\left[\; {\cal L}_{\rm F} \;+\; \frac{1}{8\pi} \left( |{\bf E}|^{2} \;-\frac{}{} |{\bf B}|^{2}
\right) \;\right] \;+\; \sum \left( \nabla{\bf u}\bdot n\,{\bf p} \;+\; \nabla\eta^{a}\;\pd{{\cal L}_{\rm F}}{\eta^{a}} \right) \;+\; \frac{1}{4\pi} \left( \nabla{\bf E}\bdot{\bf D} \;-\frac{}{} \nabla{\bf B}\bdot{\bf H}
\right), \nonumber
\end{eqnarray}
to obtain the reduced momentum conservation law \eq{momentum_t}.

\section{\label{sec:ang_mom}Reduced Angular-Momentum Conservation Law}

The symmetry properties of the reduced canonical momentum-stress tensor \eq{T_def} is physically connected to the conservation of angular momentum 
\cite{PM_85,PLS}, i.e., the exact conservation of the total angular momentum requires a symmetric {\it physical} momentum-stress tensor. Because of the obvious asymmetry of the reduced canonical momentum-stress tensor \eq{T_def}, the angular-momentum conservation law is discussed separately from the energy-momentum conservation laws discussed in the previous Section. 

According to the Noether Theorem, the component of the reduced-fluid momentum density in the direction of a spatial symmetry of the unperturbed magnetic field is conserved. In axisymmetric tokamak geometry, for example, where the background magnetic field is
\begin{equation}
{\bf B}_{0} \;\equiv\; \nabla\varphi\btimes\nabla\psi \;+\; B_{0\varphi}(\psi)\;\nabla\varphi 
\label{eq:B0_tokamak}
\end{equation}
and the background scalar fields are independent of the toroidal angle $\varphi$, we obtain
\begin{equation} 
\frac{\partial^{\prime}\ov{{\cal L}}}{\partial\varphi} \;\equiv\; \pd{{\bf x}}{\varphi}\bdot\nabla^{\prime}\ov{{\cal L}} \;=\; 0.
\label{eq:L_phi}
\end{equation}
We proceed with the derivation of the toroidal angular-momentum conservation law by following two different approaches. Both approaches identify the antisymmetric part of the reduced canonical momentum-stress tensor ${\sf T}_{\sf A} \equiv ({\sf T} - {\sf T}^{\top})/2$ as the source of intrinsic angular momentum that needs to be accounted for in expressing the exact angular-momentum conservation law. Here, the components of the antisymmetric part of the reduced canonical momentum-stress tensor \eq{T_def}, 
\begin{equation}
(T_{\sf A})_{ij} \;\equiv\; \frac{1}{2}\,\varepsilon_{ijk}\,\tau^{k},
\label{eq:T_A_ij}
\end{equation}
are expressed in terms of the reduced intrinsic torque
\begin{equation}
\vb{\tau} \;\equiv\; \sum\; n\;{\bf u}\btimes{\bf p}\;+\; {\bf E}\btimes{\bf P} \;+\frac{}{} {\bf B}\btimes{\bf M},
\label{eq:tau_ij}
\end{equation}
where we used the relations \eq{L_EB} to obtain ${\bf E}\btimes{\bf D}/4\pi = {\bf E}\btimes{\bf P}$ and $-\,{\bf B}\btimes{\bf H}/4\pi =
{\bf B}\btimes{\bf M}$. Hence, we see that the canonical momentum-stress tensor \eq{T_def} is not symmetric as a result of reduced polarization and magnetization effects and the decoupling of the reduced-fluid momentum ${\bf p} \neq m\,{\bf u}$, defined in Eq.~\eq{L_nu}, from the reduced-fluid velocity ${\bf u}$. The reduced intrinsic torque \eq{tau_ij} will now be shown to act as a source of canonical angular momentum.

\subsection{Direct approach}

By using the axisymmetry condition \eq{L_phi}, the dot product of Eq.~\eq{momentum_t} with $\partial{\bf x}/\partial\varphi$ yields the reduced toroidal-momentum transport equation
\begin{equation}
\pd{\Pi_{\varphi}^{\rm can}}{t} \;=\; -\;\pd{{\bf x}}{\varphi}\bdot\left(\nabla\bdot\frac{}{}{\sf T} \right) \;=\; -\;\nabla\bdot\left({\sf T}\bdot
\pd{{\bf x}}{\varphi} \right) \;+\; {\sf T}^{\top}:\nabla\pd{{\bf x}}{\varphi},
\label{eq:Pi_varphi}
\end{equation}
where ${\sf T}^{\top}$ denotes the transpose of ${\sf T}$ and the canonical toroidal angular-momentum density is
\begin{equation}
\Pi_{\varphi}^{\rm can} \;=\; \vb{\Pi}\bdot\pd{{\bf x}}{\varphi} \;\equiv\; \wh{\sf z}\bdot{\bf x}\btimes\vb{\Pi}.
\label{eq:Pi_varphi_def}
\end{equation} 
Using the definition \eq{pi_def} for the toroidal canonical angular momentum, we obtain
\begin{equation}
\Pi_{\varphi}^{\rm can} \;\equiv\; \sum\;n\,p_{\varphi} \;+\; \frac{{\bf D}\btimes{\bf B}}{4\pi c}\bdot\pd{{\bf x}}{\varphi},
\label{eq:Pi_phi_can_full}
\end{equation}
where $p_{\varphi} \equiv {\bf p}\bdot\partial{\bf x}/\partial\varphi$ denotes the (covariant) toroidal component of the reduced-fluid momentum. 
\
Since the mixed second-rank tensor $\nabla\partial{\bf x}/\partial\varphi \equiv \wh{R}\,\wh{\varphi} - \wh{\varphi}\,\wh{R}$ in Eq.~\eq{Pi_varphi} is antisymmetric, where $\wh{R} \equiv \nabla R = -\,\partial\wh{\varphi}/\partial\varphi$ is perpendicular to $\wh{\varphi} \equiv \wh{\sf z}\btimes
\wh{R}$, we rewrite the reduced toroidal angular-momentum transport equation \eq{Pi_varphi} as
\begin{equation}
\pd{\Pi_{\varphi}^{\rm can}}{t} \;+\; \nabla\bdot\left( {\sf T}\bdot\pd{{\bf x}}{\varphi} \right) \;=\; -\;{\sf T}_{\sf A}:\nabla\pd{{\bf x}}{\varphi} 
\;=\; -\;\left(\wh{\varphi}\btimes\wh{R}\right)\bdot\vb{\tau} \;\equiv\; \tau_{z},
\label{eq:Pi_phi_final}
\end{equation}
where we used the definition \eq{T_A_ij} and $\tau_{z}$ denotes the vertical-component of the reduced intrinsic torque \eq{tau_ij}. This equation shows that $\tau_{z}$  acts as a source of canonical toroidal angular-momentum, which can perhaps explain the spontaneous rotation of axisymmetric tokamak plasmas \cite{sp_rot_1,sp_rot_2,sp_rot_3} in the absence of external torque.

\subsection{Noether approach}

We begin our Noether approach to deriving the toroidal angular-momentum conservation law by noting that the left side of Eq.~\eq{momentum_t} is invariant under the transformation
\begin{equation}
\left. \begin{array}{rcl}
\vb{\Pi} & \equiv & \vb{\Pi}^{\rm can} \;+\; \nabla\bdot{\sf S} \\
 &  & \\
{\sf T}_{\sf S} & \equiv & {\sf T} \;-\; \partial{\sf S}/\partial t
\end{array} \right\},
\label{eq:PiT_prime}
\end{equation}
where the antisymmetric second-rank tensor ${\sf S}$ is chosen to cancel the antisymmetric part of the canonical momentum-stress tensor \eq{T_def}, i.e., ${\sf T}_{\sf A} \equiv \partial{\sf S}/\partial t$. Hence, by using the definition \eq{T_A_ij} for the reduced intrinsic torque, we obtain the definition
\begin{equation}
S_{ij} \;\equiv\; \frac{1}{2}\;\varepsilon_{ijk}\;\sigma^{k} 
\label{eq:S_sigma}
\end{equation}
for the reduced intrinsic angular momentum $\vb{\sigma}$, which satisfies the evolution equation
\begin{equation}
\pd{\vb{\sigma}}{t} \;\equiv\; \vb{\tau} \;=\; \sum\; n\;{\bf u}\btimes{\bf p}\;+\; {\bf E}\btimes{\bf P} \;+\frac{}{} {\bf B}\btimes{\bf M},
\label{eq:sigma_t}
\end{equation}
which represents a classical version of {\it zitterbewegung} \cite{Huang,Barut_Zanghi} in which dynamical reduction (e.g., the averaging of highly oscillatory motion) introduces intrinsic (spin) angular momentum.

According to the transformation \eq{PiT_prime} and the definition of the reduced intrinsic angular momentum \eq{S_sigma}, the total reduced momentum density becomes
\begin{equation}
\vb{\Pi} \;\equiv\; \vb{\Pi}^{\rm can} \;+\; \nabla\bdot{\sf S} \;=\; \vb{\Pi}^{\rm can} \;-\; \frac{1}{2}\;\nabla\btimes\vb{\sigma},
\label{eq:Pi_total}
\end{equation}
which satisfies the invariance property \cite{McLennan} $\nabla\bdot\vb{\Pi} \equiv \nabla\bdot\vb{\Pi}^{\rm can}$ and evolves according to the reduced momentum transport equation
\begin{equation}
\pd{\vb{\Pi}}{t} \;+\; \nabla\bdot{\sf T}_{\rm S} \;=\; \nabla^{\prime}\ov{\cal L}.
\label{eq:total_momentum}
\end{equation}
The toroidal the reduced toroidal angular-momentum conservation now becomes
\begin{equation}
\pd{\Pi_{\varphi}}{t} \;+\; \nabla\bdot\left({\sf T}_{\sf S}\bdot\pd{{\bf x}}{\varphi} \right) \;=\; 0,
\label{eq:Pi_phi_t}
\end{equation}
where the total reduced toroidal angular-momentum density is
\begin{equation}
\Pi_{\varphi} \;=\; \vb{\Pi}\bdot\pd{{\bf x}}{\varphi} \;=\; \Pi_{\varphi}^{\rm can} \;-\; \sigma_{z} \;+\; \nabla\bdot\left({\sf S}\bdot\pd{{\bf x}}{\varphi}\right),
\label{eq:Pi_phi_phys}
\end{equation}
where we used ${\sf S}:\nabla\partial{\bf x}/\partial\varphi = -\,\sigma_{z}$. By substituting Eq.~\eq{Pi_phi_phys} into Eq.~\eq{total_momentum}, we recover Eq.~\eq{Pi_phi_final} with $\tau_{z} \equiv \partial\sigma_{z}/\partial t$.

\subsection{Toroidal angular-momentum conservation in axisymmetric tokamak geometry}

Another useful form of the reduced toroidal-momentum transport equation \eq{Pi_phi_final} [or Eq.~\eq{Pi_phi_t}] is expressed as
\begin{equation}
\pd{\langle\Pi_{\varphi}^{\rm can}\rangle}{t} \;+\; \frac{1}{{\cal V}}\;\pd{}{\psi} \left( {\cal V}\frac{}{}\left\langle T_{\varphi}^{\psi} 
\right\rangle \right) \;=\; \langle\tau_{z}\rangle \;\equiv\; \pd{\langle\sigma_{z}\rangle}{t},
\label{eq:Pi_toroidal_ave}
\end{equation}
where $\langle\cdots\rangle \equiv {\cal V}^{-1}\,\oint (\cdots)\,{\cal J}\,d\vartheta\,d\varphi$ denotes the magnetic-surface average (labeled by the magnetic flux $\psi$), with ${\cal J} \equiv (\nabla\psi\btimes\nabla\vartheta\bdot\nabla\varphi)^{-1}$ denoting the Jacobian associated with the magnetic coordinates $(\psi, \vartheta, \varphi)$ and ${\cal V} \equiv \oint{\cal J}\,d\vartheta\,d\varphi$. In Eq.~\eq{Pi_toroidal_ave}, the surface-averaged toroidal angular-momentum flux is
\begin{equation}
T^{\psi}_{\;\;\varphi} \;\equiv\; \nabla\psi\bdot{\sf T}\bdot\pd{{\bf x}}{\varphi} \;=\; \sum\,n u^{\psi}\,p_{\varphi} \;-\; \frac{1}{4\pi}\,\left( 
D^{\psi}\,E_{\varphi} \;+\frac{}{} B_{\bot}^{\psi}H_{\varphi}\right),
\label{eq:T_psi_varphi}
\end{equation}
where $B_{\bot}^{\psi} \equiv {\bf B}_{\bot}\bdot\nabla\psi$ denotes the $\psi$-component of the perpendicular component of the perturbed magnetic field (since ${\bf B}_{0}\bdot\nabla\psi \equiv 0$). 

In the electrostatic limit $({\bf B} \equiv {\bf B}_{0}, {\bf E} \equiv -\,\nabla\Phi)$ and making use of the quasi-neutrality condition $D^{\psi} 
\simeq 4\pi\,P^{\psi}$, the canonical toroidal angular momentum \eq{Pi_phi_can_full} becomes
\[ \Pi_{\varphi}^{\rm can} \;=\; \sum\; n\,p_{\varphi} \;+\; \frac{1}{c}\;P^{\psi}, \]
where we used the identity ${\bf B}_{0}\btimes\partial{\bf x}/\partial\varphi \equiv \nabla\psi$, and the toroidal angular-momentum flux 
\eq{T_psi_varphi} becomes
\[ T^{\psi}_{\;\;\varphi} \;=\; \sum\,n\; u^{\psi}\,p_{\varphi} \;+\; P^{\psi}\;\pd{\Phi}{\varphi}. \]
Hence, in the electrostatic limit, the surface-averaged toroidal angular-momentum transport equation \eq{Pi_toroidal_ave} becomes
\begin{equation}
\pd{}{t} \left( \sum\; \langle n\,p_{\varphi}\rangle \;+\; \frac{1}{c}\;\langle P^{\psi}\rangle \right) \;+\; \frac{1}{{\cal V}}\;\pd{}{\psi} \left[ 
{\cal V}\;\left( \sum\,\left\langle n u^{\psi}\,p_{\varphi}\right\rangle \;+\; \left\langle P^{\psi}\;\pd{\Phi}{\varphi}\right\rangle \right) \right]
\;=\; \langle \tau_{z}\rangle,
\label{eq:Pi_electrostat}
\end{equation}
where the surface-averaged intrinsic torque involves the reduced intrinsic torque component
\begin{eqnarray}
\tau_{z} & = & \wh{\sf z}\bdot\left[\;\sum\;n\;{\bf u}\btimes{\bf p} \;+\; \left( {\bf E}\btimes{\bf P} \;+\; {\bf B}_{0}\btimes{\bf M}
\right) \;\right] \nonumber \\
 & \equiv & \sum\;mn\,u_{\|}\;\wh{\sf z}\bdot\left[\left({\bf u}_{\bot} \;-\frac{}{} {\bf U}_{\bot}\right)\btimes\bhat_{0} \right].
\label{eq:tau_z_electrostat}
\end{eqnarray}
The last expression for $\tau_{z}$ in Eq.~\eq{tau_z_electrostat} makes use of the gyrofluid relations \cite{Brizard_NFLR} ${\bf p} \equiv m\,u_{\|}\,\bhat_{0}$, ${\bf E}\btimes{\bf P} \equiv 0$ (in the electrostatic limit), and ${\bf B}_{0}\btimes{\bf M} \equiv \sum\,(mn\,u_{\|})\,\bhat_{0}\btimes
{\bf U}_{\bot}$, where ${\bf U}_{\bot}$ denotes the perturbed $E\times B$ velocity (i.e., the lowest-order perturbed perpendicular reduced-fluid velocity). Equation \eq{Pi_electrostat} was recently derived by Scott and Smirnov \cite{Scott_Smirnov}, without the reduced intrinsic torque, by considering the time evolution of the gyrocenter moment of the (guiding-center) canonical toroidal angular momentum $mv_{\|}\,\bhat_{0} - (q/c)\,\psi$. 

Lastly, we note from Eq.~\eq{tau_z_electrostat} that, since the perpendicular reduced velocities ${\bf u}_{\bot}$ and ${\bf U}_{\bot}$ are typically driven by perpendicular gradients of plasma parameters \cite{Brizard_NFLR} (e.g., electrostatic potential and plasma pressure), the vertical component of the reduced intrinsic torque \eq{tau_z_electrostat} appears to predominantly involve the vertical components of these perpendicular plasma-parameter gradients. Hence, we expect that the surface-averaged reduced intrinsic torque $\langle\tau_{z}\rangle$ might be greatly determined by the up-down symmetry of the axisymmetric magnetized plasma \cite{sp_rot_11}.

\section{\label{sec:summary}Summary}

In the present paper, we have used a variational principle based on the general reduced Lagrangian density \eq{L_F_def} to derive general reduced-fluid dynamical equations \eq{Phi_EP}-\eq{u_EP} subject to the constraint equations \eq{Maxwell_00}-\eq{pper_eq}. These reduced equations satisfy the energy-momentum conservation laws \eq{energy_t} and \eq{momentum_t} as well as the toroidal angular-momentum conservation law \eq{Pi_phi_t}.

The case of the toroidal angular-momentum conservation law \eq{Pi_phi_t} is particularly interesting because it identifies a reduced intrinsic torque $\vb{\tau}$ generated by the dynamical reduction, which generates polarization and magnetization effects as well as a decoupling between the reduced-fluid momentum and the reduced-fluid velocity. The vertical component $\tau_{z} \equiv \partial\sigma_{z}/\partial t$ of the reduced intrinsic torque was shown in Sec.~\ref{sec:ang_mom} to be the source of intrinsic toroidal angular-momentum $\sigma_{z}$, which can generate spontaneous toroidal rotation in the absence of external torque.

Future work will focus on the physical interpretation of the reduced intrinsic torque \eq{tau_ij}, and the reduced intrinsic angular momentum 
$\vb{\sigma}$ it generates, as well as investigating the surface-averaged intrinsic torque $\langle\tau_{z}\rangle$ for several reduced-fluid models \cite{Brizard_PRL}. Ongoing work (with N.~Tronko) is also investigating the reduced toroidal angular-momentum conservation law for the nonlinear gyrokinetic Vlasov-Maxwell equations \cite{Brizard_Hahm}.

\acknowledgments

The Author benefited from insightful discussions with John A.~Krommes while at the Isaac Newton Institute for Mathematical Sciences. The Author also thanks Phil Morrison for information concerning the Belinfante derivation of a symmetric stress-momentum tensor. This work was supported by a U.~S.~Department of Energy grant No.~DE-FG02-09ER55005.


\begin{thebibliography}{99}
\bibitem{footnote} The notation $\vb{\Pi}$ is used for the reduced momentum density (and ${\sf T}$ is used for the reduced momentum-stress tensor) in order to avoid using the notation ${\bf p}$ for the reduced kinetic momentum \eq{L_nu}, the notation ${\bf P}$ for the reduced polarization 
\eq{pol_mag}, or the notation ${\sf P}$ for the CGL pressure tensor \eq{P_CGL}.
\end{thebibliography}
\end{document}